\documentclass[epj]{svjour}
\usepackage{graphicx}
\begin{document}
\title{Frozen capillary waves on glass surfaces: an AFM study}
\author{Thomas Sarlat \and Anne Lelarge \and Elin S{\o}nderg{\aa}rd
\and Damien Vandembroucq\thanks{corresponding author: 
damien.vdb@saint-gobain.com}
}                     
\offprints{}          
\institute{Laboratoire ``Surface du Verre et Interfaces'' \\
Unit\'{e} Mixte CNRS/Saint-Gobain  \\
39 quai Lucien Lefranc, F-93303 Aubervilliers Cedex, France.}
\date{Received: date / Revised version: date}
%
\abstract{Using atomic force microscopy on silica and float glass
surfaces, we give evidence that the roughness of melted glass surfaces
can be quantitatively accounted for by frozen capillary waves. In this
framework the height spatial correlations are shown to obey a
logarithmic scaling law; the identification of this behaviour allows to
estimate the ratio $kT_F/\pi\gamma$ where $k$ is the Boltzmann
constant, $\gamma$ the interface tension and $T_F$ the temperature
corresponding to the ``freezing'' of the capillary waves. Variations
of interface tension and (to a lesser extent) temperatures of
annealing treatments are shown to be directly measurable from a
statistical analysis of the roughness spectrum of the glass surfaces.
}
\maketitle
\section{Introduction}
\label{intro}


Glass is becoming a widespread substrate in high-tech developments
like flat displays \cite{E1}, thin foiled x-ray telescope mirrors \cite{E2} and
lithography masks \cite{E3}. In all of these applications, controlling
the glass surface roughness is a critical issue to enhance
performance. Glass manufactures spend considerable effort in
optimizing their forming process.  The introduction in the late
fifties of the float process to produce flat glass represented a
revolution in the glass industry: the glass sheets obtained in this
way are almost perfectly flat so that the expensive polishing step
revealed to be unnecessary for most applications. The process somehow mimics
and extends the traditional method of fire polishing. The latter
consists of first heating the surface to a temperature high enough to
obtain local melting and then quenching the resulting smooth surface.
Laser polishing \cite{Laser-polishing} can be seen as another modern
extension of the same strategy. In the float process, the molten glass
flows from the furnace on a liquid tin bath. The
(liquid) glass thus spreads out on the tin surface to form a ribbon. The
progressive cooling imposed along the tin bath
allows the two glass surfaces (tin/glass and glass/atmosphere) to
freeze without external mechanical contact. Note that similar comments could be made about other glass
forming processes like fiber drawing.


The residual roughness of these ``fire polished'' surfaces is
extremely low, typically below the nanometer range when measured over
a micrometric scale and has been attributed to the freezing of the
capillary waves of the liquid interface in the glass transition
temperature range\cite{Jackle-JPC95}. This particular phenomenon has
recently been studied on glycerol\cite{Seydel-PRB02,Madsen-PRL04} and
supercooled polymers\cite{Sprung-PRB04} by X-ray reflectivity. Though
the surface roughness induced by capillary waves is long range
correlated, the X-ray reflectivity of a liquid surface was shown to
depend mainly on the total interface width $\sigma\propto kT/\gamma$
\cite{Ocko-PRL94}, where $k$ is the Boltzmann constant, $T$ the
temperature and $\gamma$ the interface tension. When varying
temperature in the glass transition regime, a clear saturation of the
interface roughness was obtained below a threshold
temperature\cite{Seydel-PRB02,Madsen-PRL04,Sprung-PRB04}. A
satisfactory quantitative agreement with a capillary waves description
is obtained in the case of polymers; for glycerol, however, a
density driven layering effect was suspected, only a qualitative
agreement could be obtained.


In this paper, we present a detailed quantitative atomic force
microscopic (AFM) study of the roughness spectrum of melted glass
surfaces, namely silica surfaces obtained after various annealing
treatments and bare industrial float glass surfaces.  AFM has
previously been used to study the roughness of glass
surfaces\cite{Abriou-96};
but only recently, a few reports \cite{Gupta-JNCS00,Roberts-OE05} have
been given of a qualitative agreement between the values of the
roughness measured on glass surfaces and the simple estimates obtained
from a description of the surface in terms of frozen capillary waves
$\sigma \approx kT_G/\gamma$. Here again,  $\sigma$ is the standard deviation
of the height fluctuations along the surface (RMS roughness), $\gamma$
the surface tension of the glass and the glass transition temperature
$T_G$ is used as an estimate of the freezing temperature of capillary
waves. Focusing on two different glass surfaces (silica and float
glass) we show here that the height correlations along the surface
give a direct and quantitative access to the frozen capillary wave
spatial spectrum. The use of these two different materials allows
us study two different aspects of the phenomenon. Performing
heat treatments on silica at different temperatures in the glass
transition range we can study the effect of thermal history on the
glass surface roughness. Performing measurements on the two faces
(namely ``tin'' and ``atmosphere'') of a float glass, we can test the
effect of the two different interface tensions due to the asymmetry of
the float process.

The paper is organised as follows. We first  recall some results on
 capillary waves at a liquid interface and we give
quantitative expressions expected for the height correlations in this
framework. Then we detail the experimental protocol and we finally
present the AFM measurement results obtained respectively on silica
and float glass. Finally we discuss their interpretation in terms of
frozen capillary waves.

\section{Capillary waves}

Assuming a constant value of the surface tension $\gamma$, the excess
of energy $E(h)$ due to the height fluctuations $h(x_1,x_2)$ of a
liquid interface can be estimated to:

\begin{eqnarray}
E(h)&=&\gamma \int\left(\sqrt{1+|\nabla h^2|}-1\right)dx_1dx_2 \;,\\
&\simeq &\frac{\gamma}{2} \int|\nabla h|^2 dx_1dx_2  \;,
\end{eqnarray}
\noindent where $x_1,x_2$ are the horizontal coordinates
 and
$\gamma$ the interface tension. Using a Boltzman
distribution, we easily obtain the expressions of the RMS roughness  $\sigma$ and of the spatial correlation function $G(r)$ and
of the height variogram $H(r)$:
\begin{equation}
\label{eqn:001}
\sigma^2=\langle h^2(\overrightarrow{x})\rangle_{\overrightarrow{x}}
=\frac{kT}{2\pi\gamma}\ln(\frac{\lambda_M}{\lambda_{m}})
\end{equation}
\begin{equation}
G(r)=\langle h(\overrightarrow{x}+\overrightarrow{r})
h(\overrightarrow{x})\rangle_{\overrightarrow{x}}=\frac{kT}{2\pi\gamma}
\ln(\frac{\lambda_M}{r})
\label{eqn:002}
\end{equation}
\begin{equation}
H(r)=\langle |h(\overrightarrow{x}+\overrightarrow{r})
-h(\overrightarrow{x})|^2 \rangle_{\overrightarrow{x}}
=\frac{kT}{\pi\gamma}\ln(\frac{r}{\lambda_{m}})
\label{eqn:002b}
\end{equation}

where
$\overrightarrow{x}=x_1\overrightarrow{e_1}+x_2\overrightarrow{e}_1$
and $r=|\overrightarrow{r}|$. As the expressions are
logarithmically divergent,  an upper
($\lambda_M$) and a lower ($\lambda_m$) spatial cut-off 
\ must be introduced. Gravity
effects are neglected in the above description and the upper
cut-off is given by the capillary length
$\ell_c=\sqrt{\gamma/\rho g}$ where $\rho$ is the glass density
and $g$ is the gravity constant. For silica ($\gamma\simeq0.3
\mathrm{J.m}^{-2}$, $\rho \simeq 2.2 \mathrm {kg.m}^{-3}$) this
length is approximatively equal to 4 mm. The lower cut-off is
 estimated  to a molecular length $\ell_0$. For silica
and silicate glasses in general the structural unit is the
SiO$_{4}$ tetrahedron, and we can take $\ell_0\simeq 0.5 {\mathrm
nm}$. The typical roughness of a silica surface in the glass
transition region ($T_G \simeq 1450$) due to capillary
waves can thus be estimated to $\sigma \approx 1.5 \mathrm{nm}$.
In a first approximation we consider that the interface height
fluctuations freeze at this glass transition temperature $T_G$ so
that equations (\ref{eqn:001},\ref{eqn:002},\ref{eqn:002b}) can be
used to describe the roughness of the glass surfaces at room
temperature simply replacing $T$ by $T_G$. A detailed treatment of
the freezing of capillary waves based on the visco-elasticity
theory can be found in \cite{Jackle-JPC95}.

Molecular and capillary length thus define the ultimate physical
bounds of the expected logarithmic scaling. However, in the
framework of roughness measurement, the cut-off lengths are imposed
by the limited bandwidth of the experimental set-up. In the context of
atomic force microscopy the scan size, usually in the micromemeter
range, defines the upper cut-off $\lambda_M$ and the nanometric
tip radius (or the sampling length) define the lower cut-off
$\lambda_m$.

\section{Experimental method}
We detail here  the sample preparation (annealing treatment for
silica and cleaning procedure) and on the roughness data
extraction from the raw AFM data.

\subsection*{Sample preparation}

\subsubsection*{Float glass}
Float glass  is a standard industrial product 
with approximate composition: SiO$_2$ $ 72\%$, Na$_2$O $14 \%$,
CaO $9 \%$, MgO $4 \%$. Note that due to the asymmetry of the
float process some traces of Sn are present in the vicinity of the
``tin'' surface. 1 cm$^2$ samples were cut from a 3 mm
 thick glass sheet. No additional treatment but
cleaning (described below) was performed on these surfaces.

\subsubsection*{Silica}
The silica samples are commercial silica wafers (vitreosil, EQ512), 2
mm-thick, two side polished. Square samples of 1cm$\times$1cm were
prepared. Prior to the annealing treatment, the sample preparation was
as follows: the sample is supported on a finely grained silicon
carbide plate, Saint-Gobain Ceramics, and covered by a platinum lid, the entire system is placed on a fire-brick. Fire-brick,
platinum lid, silicon carbide plate and samples are cleaned with the
same protocol : i) one ultra-sonic cleaning of fifteen minutes in a
millipore water/detergent (aquanox) 10\% mixture, ii) two ultra-sonic
rinsing of fifteen minutes in millipore water, iii) drying with dry
nitrogen. After this cleaning, all pieces are  placed in a
furnace at 60$^\circ$C to dry the porous fire-brick.

\subsubsection*{Annealing treatments on silica}

All silica samples were first submitted to a high temperature
treatment: 1770 K during 30 minutes.  This step is performed
well above the $T_G\simeq 1450{\mathrm K}$ to ensure full
relaxation of surface defects due to the polishing step.  The sample
is then immediately taken out from the furnace; the platinum cap is
only removed after 10 minutes  in order to avoid dust in-layering on
the molten surface

Beyond this surface relaxation step, the
silica samples have undergone different thermal treatments  to
test the dependence of surface roughness on thermal history. We
followed annealing protocols proposed  by Le Parc {\it et al.}
\cite{LeParc-JNCS01}  who studied the effect of thermal history on
the Raman spectrum of silica.
The thermal treatments are listed in table
\ref{tab:001}. Annealing times have been calculated to allow silica
relaxation. After annealing each sample is removed from the furnace in
the same condition as described above.

\begin{table}[htbp]
\begin{center}
\begin{tabular}{|l|c|c|c|c|}
\hline
Sample & 1500 & 1350 & 1200 & 1100  \\
\hline
A   & 30 min & & & \\
B   & 30 min & 60 min & & \\
C   & 30 min & & 120 min & \\
D   & 30 min & & & 68 hours \\
\hline
\end{tabular}
\end{center}
\caption{Thermal history of samples}
\label{tab:001}
\end{table}

Note that the long annealing time applied to sample D has caused a
slight nucleation of a crystalline phase of silica under the
surface. This is not expected to affect further interpretations as
the crystals were subsurface.

\subsubsection*{Post-annealing cleaning}

In order to eliminate contaminations due to storage and transport,
samples were cleaned before AFM measurements. We checked that the
surface was not altered by successive cleaning operations.  The
previous cleaning protocol was used. A last step is added : a 45
minutes UV/O$_{3}$ cleaning to remove nanometric dust and also to
obtain a hydrophilic surface improving surface imaging. Despite theses
precautions; some dust was found on the surfaces . However, AFM scans
show large dust free parts (typically 15$\mu$m*15$\mu$m).

\begin{figure}[phtb]
\begin{center}
\includegraphics[width=0.37\textwidth]{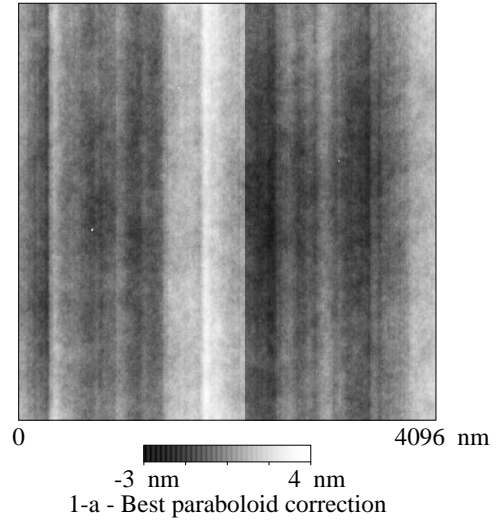}
\includegraphics[width=0.37\textwidth]{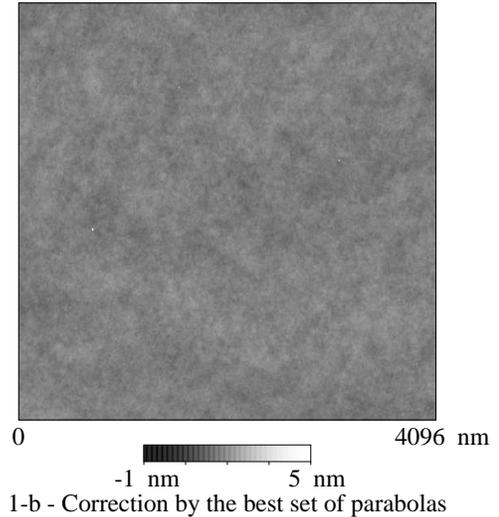}
\caption{An image of an annealed silica surface (A) acquired in the
open air. a) correction by the best fitted paraboloid
($\sigma_{rms}=0.43\ nm$) to compensate the systematic curvature
effect due to the spherical motion of the piezo-electric stage, note
the residual bands due to a drift error of the tip; b) correction by
the best set of parabolas ($\sigma_{rms}=0.25\ nm$). }\label{fig:01}
\end{center}
\end{figure}

\subsection*{Quantitative use of AFM for surface morphology.}

The surface morphology of all samples have been characterized by AFM
height measurement in tapping mode (TM-AFM), using a Nanoscope III A
from Digital Instruments with Al coated tip (BudgetSensors - model
BS-Tap 300 Al). The Al coating thickness is 30nm, the resonant
frequency is 300 kHz and the stiffness constant is 40N/m. The tip
radius is lower than 10 nm. The images have been recorded at a scan
frequency between 0.8 and 1Hz for a resolution of $512\times512$
pixels.

As our study is focused on the analysis of extremely smooth surfaces, special attention has to be paid to the data treatment. The
roughness induced by capillary waves lies indeed in the sub-nanometer
range and corresponds to the ultimate thermodynamic fluctuations of
the interface. At this low roughness level, limitations of AFM can be
observed in the bare data.  

The sample is moved by a XYZ-piezo-electric stage and long profiles
exhibit a slight spherical character. This systematic curvature in the
height data can be eliminated by subtracting the best paraboloid to
the surface. Fig. \ref{fig:01}\---a) shows the result of this first
operation. A remaining strong band pattern is observed; this results
from a drift error of re-positioning of the piezo-electric
stage. Strictly speaking AFM measurements thus have to regarded as a
set of profiles more than an image and bare data must be corrected
line by line. Fig.  \ref{fig:01}\---b) shows a representative AFM
image obtained after subtraction of the best set of parabolas (in a
least mean square sense) to the set of profiles. The statistical
analysis is thus performed on these profiles.

\section{Results}

\begin{figure}[b]
\begin{center}
\vspace{0.3cm}
\includegraphics[width=0.39\textwidth]{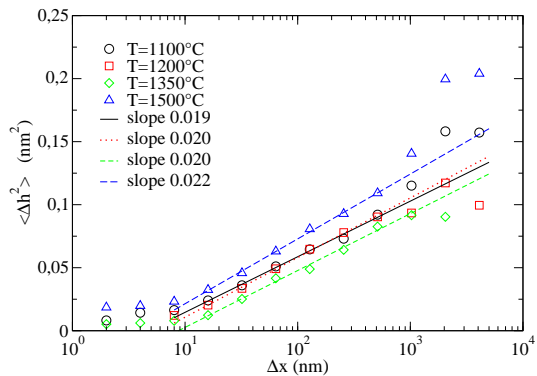}
\caption{Height variogram obtained from AFM measurements silica
surfaces after various annealing treatments in the glass transition
regime (symbols). The lines correspond to the predicted
logarithmic behaviour induced by the presence of frozen capillary
waves (fit performed in the interval [16nm-500nm])}\label{fig:03}
\end{center}
\end{figure}

\begin{figure}[b]
\begin{center}
\vspace{0.3cm}
\includegraphics[width=0.39\textwidth]{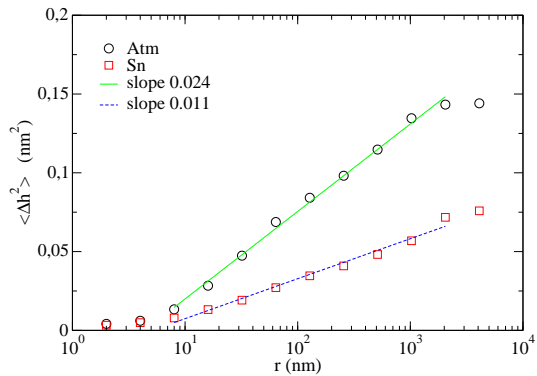}
\caption{Height variogram obtained from AFM measurements on the
two faces of a float glass, respectively in contact with liquid
tin and with atmosphere (symbols). The fitted lines correspond to the
logarithmic  behaviour predicted by the presence of frozen capillary
waves (fit performed in the interval [8nm-2000nm]).}\label{fig:04}
\end{center}
\end{figure}

\subsection{Silica}

A set of AFM roughness measurements on the samples A,B,D has been made in
order to determine the scaling behaviour of the roughness from the
nanometer range up to the micrometer range. The typical experiment is
the following : a first image is acquired over 8$\mu$m$\times$8$\mu$m,
then a hierarchy of images of decreasing size (4$\mu$m,
2$\mu$m... down to 512 or 128 nm depending on the sample) is taken
within this first image. We performed the measurements of 2 such
series down to 128nm for sample A; 2 series down to 512nm for sample B
and 3 series down to 512nm for sample D. On sample C we measured 14
images of 2$\mu$m$\times$2$\mu$m and 7 sets of 2 imbricated images of
respective size 8$\mu$m$\times$8$\mu$m and 2$\mu$m$\times$2$\mu$m.

The statistical treatment consisted in computing the height variogram
on each individual image and to average these data for the different
images of a series. We thus obtain on Fig. \ref{fig:03}, the evolution
of the average quadratic height difference $H(r)=\langle |h({x}+{r})
-h({x})|^2 \rangle_{{x}}$ between points on the surface separated by a
distance $r$. In the semi-log scale used in Fig. \ref{fig:03}, we
obtain a linear behaviour over more than 2 decades from a few nanometers
up to a few micrometers.  We thus obtain a very good agreement with the
predicted capillary wave scaling proposed in Eq. (\ref{eqn:001}). The
values of the slopes obtained by a simple linear fit in the scaling
region are reported in Table \ref{tab:002}.  A physical discussion on
these extracted parameters and the quantitative agreement with the
theory is given in section \ref{discu}.

\subsection{Float glass}

Using the same procedure as above, 6 sets of images of decreasing
size from 8$\mu$m$\times$8$\mu$m down to 500nm$\times$500nm were
performed on the atmosphere and tin faces of the float glass
samples. The atmosphere  faces were more affected by
residual dust pollution and only 3 of them could be quantitatively
analysed. 
The height
variograms are displayed in semi-log scale in Fig. \ref{fig:04}.
Again we observe a logarithmic behaviour
over more than two decades. The values of the slopes obtained by a
simple linear fit in the scaling region are reported in Table
\ref{tab:002}. The tin face appears  to be much
smoother than the atmosphere face, which is consistent with the
smaller interface tension expected for the glass atmosphere
interface compared with the tin/glass interface.

\subsection{Statistical averaging, uncertainty}
To evaluate the accuracy of our results, we investigated the
dispersion of the roughness measurements. This was done on the silica
sample C. As a first test, 14 images of size 2 $\mu$m were recorded
within a primary zone of size 8 $\mu$m. The mean value of the height
variances obtained on these images was $\sigma^2=0.029 \mathrm{nm}^2$
and the standard deviation 0.005 nm$^2$, that is to say 15 \% of the
mean value. As a second test, we performed 7 sets of 2 imbricated
images of respective sizes 8 $\mu$m and 2$\mu$m measured at different
places over the surface.  For the 8 $\mu$m scans, we obtain a mean
value of 0.053 nm$^2$ and a root mean square of 0.007 nm$^2$ that is
to say 15 \% of the mean value. For the 2 $\mu$m we found a mean value
of 0.030 nm$^2$ and a root mean square of 12 \% of the mean value. We
can conclude that we have roughly a dispersion of 15 \% of the mean
value.  This gives us an estimate of the uncertainty on our roughness
measurements and on the numerical parameters used to fit the
logarithmic behaviour predicted by Eq. (\ref{eqn:002b}).

\begin{table}[t]
\begin{center}
\begin{tabular}{|l|c|c|c|c|}
\hline
Sample &T$_F$(K) &$\gamma$ (J/m$^{2}$) & $a_{th}$ (nm$^2$)
&a (nm$^2$)  \\
\hline
silica A &1773  &0.3   & 0.026  & 0.022$\pm$0.003\\
silica B &1623  &0.3   & 0.024  & 0.020$\pm$0.003\\
silica C &1473  &0.3   & 0.022  & 0.020$\pm$0.003\\
silica D &1373  &0.3   & 0.020  & 0.019$\pm$0.003\\
float atm. &873 &0.35  & 0.011  & 0.024$\pm$0.004\\
float tin  &873 &0.5   & 0.008  & 0.011$\pm$0.002\\
\hline
\end{tabular}
\end{center}
\caption{Estimation of the capillary wave slope parameter obtained by
fitting the evolution of the mean square : $H(r)=a r+b$ and comparison
with reference values $a_{th}=\frac{kT_{F}}{\pi\gamma}$ obtained from
the annealing temperatures and estimates of the interface tensions.}
\label{tab:002}
\end{table}

\section{Discussion}
\label{discu}

As shown in Fig.\ref{fig:03} and \ref{fig:04} the height variograms
show a clear logarithmic behaviour from the nanometer range up to the
micrometric range. Table \ref{tab:002} shows a summary of the parameters
characterising the frozen capillary wave regime. As references,
we indicated the values ${kT_{F}}/{\pi\gamma}$ corresponding
to the annealing temperatures used for silica and the temperature
of the soda-lime glass at the end of the float tank; we used
generally accepted estimates of the interface tensions silica/air,
glass/air and tin/glass at high
temperature\cite{Scholze,Barton,Nizhenko03}.

Several comments can be made upon these results. First, within
experimental errors we obtain a clear logarithmic scaling of the
spatial correlation with physical parameters consistent with the
scenario of capillary waves freezing.

In this naive scenario we assume that the surface freezes at a
well-defined temperature $T_F$.  Some more complex phenomena may
characterise the freezing step, in particular one can think of a
dependence of the freezing temperature on the spatial frequency
dependence of the capillary waves. However such glassy behaviours,
leading to departures from the simple logarithmic scaling could not be
identified within experimental errors.

The logarithmic behaviour could be observed over 2 to 3 decades from
the nanometer range to the micrometer range. Our data do not show any
clear upper cut-off. Our AFM set-up being limited to a maximum scan
length of 17 $\mu$m, larger scales were not explored. Note however
that the pollution of the surfaces by dust or impurities makes it more
and more difficult to analyse the data beyond 10 $\mu$m.  As discussed
above, the lower cut-off of the logarithmic range is supposed to
correspond to an atomic length scale. In our case, this cut-off has an
instrumental origin and amounts to a few nanometers, which correponds
to the AFM tip resolution. Note again that the logarithmic scaling
induces a divergence for both small and large scales. Variations of
the lower cut-off value due to a change of rounding of the AFM tip
thus may induce a vertical shift of the roughness data (see
Eq. (\ref{eqn:002b})).

In this framework of frozen capillary waves, the slope measured in the
semi-log plot is directly related to the ratio $T_F/\gamma$ where
$T_F$ is the freezing temperature and $\gamma$ the interface tension.

In the case of silica we observe first a very good agreement between
experimental slopes and references values. We observe a very low
growing tendency of this slope with the annealing temperature. This
may thus correspond to a slight increase of the effective freezing
temperature but this tendency is certainly not clear enough to give
any firm conclusion. The typical slope obtained around 0.20 nm$^2$ are
very close to the value 0.22 nm$^2$ corresponding to the glass
transition temperature of silica $T_G=1450 \mathrm{K}$ and an
interface tension $\gamma=0.3 \mathrm{J.m}^{-2}$. Let us note in
addition that, because of thermal inertia, it is not surprising that
for the annealing processes at high temperature the effective freezing
temperature (the experimental slope) be lower than the annealing
temperature (the reference slope).

In the case of float glass we observe a striking contrast between the
two faces. The tin face is noticeably smoother than the atmosphere
face as can be expected from the contrast of interface
tensions. Focusing on the quantitative aspect, we note that the
experimental slopes are quite higher than the reference values,
particularly in the case of the atmosphere face. This may obviously be
due to an underestimation of the freezing temperature and/or an
overestimation of the interface tensions. Let us insist on the latter
point, the experimental measurement of interface tensions at high
temperature (sessile drop methods and further evolutions) remains
delicate and large variations of the measured values can be found in
literature\cite{Scholze}. The interface tension can indeed be very
dependent on the atmosphere and on the composition.

\section{Conclusion}

Performing AFM roughness measurements on annealed silica surfaces and
on float glass surfaces we could confirm that the roughness of ``fire
polished'' glasses can be quantitatively described by frozen capillary
waves in the micrometric range. In particular we could identify a
clear dependence of the roughness spectrum on the interface tension
and we obtain indications of a slight dependence upon the freezing
temperature.  At length scales below the capillary length, the
industrial float process thus allows to produce flat glass surfaces
characterised by the lowest thermodynamically possible roughness; the
only way of reducing these fluctuations would consist of increasing the
interface tension or of decreasing the ``freezing'' temperature of
capillary waves. In addition, we could show that the description of
glass surfaces by capillary waves gives access to the ratio freezing
temperature on interface tension $T_F/\gamma$ and can thus be
considered as a complementary indirect way of measuring the two
quantities.

\section*{Acnowledgements}
We acknowledge useful discussions with D. Dalmas, P. Garnier and
S. Roux.  We are grateful for financial assistance from the EU
FP6 NEST project ``Imaging by Neutral Atoms''.

%
%

\end{document}